\documentclass[conference]{IEEEtran}
\usepackage[
draft,
pdfpagelabels=false,
pdfpagelabels=false,
hyperindex=false,
pageanchor=false
]{hyperref}
\usepackage{setspace}	
\usepackage{latexsym}
\usepackage{multirow}
\usepackage{placeins}
\usepackage{caption}

\usepackage{subcaption}
\usepackage{float}
\usepackage{amsmath}
\usepackage{algorithm}
\usepackage{enumitem}
\setlist{nolistsep}
\usepackage{graphics,graphicx}
\usepackage{comment}
\usepackage{color}
\usepackage{multirow}
\usepackage{adjustbox}
\usepackage{hhline}

\begin{document}
	%
	% paper title
	% Titles are generally capitalized except for words such as a, an, and, as,
	% at, but, by, for, in, nor, of, on, or, the, to and up, which are usually
	% not capitalized unless they are the first or last word of the title.
	% Linebreaks \\ can be used within to get better formatting as desired.
	% Do not put math or special symbols in the title.
	\title{Understanding Psycholinguistic Behavior of predominant drunk texters in Social Media}

	% author names and affiliations
	% use a multiple column layout for up to three different
	% affiliations
	\author{\IEEEauthorblockN{Suman Kalyan Maity$^1$\IEEEauthorrefmark{1}, Ankan Mullick$^2$\IEEEauthorrefmark{1}, Surjya Ghosh$^3$, Anil Kumar$^4$, Sunny Dhamnani$^5$, Sudhansu Bahety$^6$ \\ and  Animesh Mukherjee$^3$}
\IEEEauthorblockA{Northwestern University$^1$; Microsoft, India$^2$; IIT Kharagpur, India$^3$; Samsung, India$^4$; Adobe Research, India$^5$; Salesforce, USA$^6$ \\
Email: suman.maity@kellogg.northwestern.edu$^1$, ankan.mullick@microsoft.com$^2$ \\
\IEEEauthorrefmark{1}Authors have contributed equally. Most of the work was done when all the authors were in IIT Kharagpur, India.}}
%; Adobe Research, Bangalore, India$^4$; Salesforce, USA$^5$	

	% make the title area
	\maketitle
	
	% As a general rule, do not put math, special symbols or citations
	% in the abstract
	\begin{abstract}
		In the last decade, social media has evolved as one of the leading platform to create, share, or exchange information; it is commonly used as a way for individuals to maintain social connections. In this online digital world, people use to post texts or pictures to express their views socially and create user-user engagement through discussions and conversations. Thus, social media has established itself to bear signals relating to human behavior. One can easily design user characteristic network by scraping through someone's social media profiles. In this paper, we investigate the potential of social media in characterizing and understanding predominant drunk texters from the perspective of their social, psychological and linguistic behavior as evident from the content generated by them. Our research aims to analyze the behavior of drunk texters on social media and to contrast this with non-drunk texters. We use Twitter social media to obtain the set of drunk texters and non-drunk texters and show that we can classify users into these two respective sets using various psycholinguistic features with an overall average accuracy of 96.78\% with very high precision and recall. Note that such an automatic classification can have far-reaching impact -- (i) on health research related to addiction prevention and control, and (ii) in eliminating abusive and vulgar contents from Twitter, borne by the tweets of drunk texters.
	\end{abstract}
	%\keywords{Drunk-texters, psycholinguistic behavior, computational health}
	%\maketitle	
	\IEEEpeerreviewmaketitle

	\section{Introduction}
	Alcohol consumption has serious implications on individual's health. In 2012, 5.9\% of all global deaths (7.6\% for men and 4.0\% for women), were attributed to alcohol consumption and the number is increasing over time. In US alone, nearly 88,000 people (approximately 62,000 men and 26,000 women) die from alcohol-related causes yearly, making it the fourth leading preventable cause of death in that country\footnote{http://1.usa.gov/1hcR6dX}. In addition to causing traumatic death and injury, alcohol consumption also leads to chronic liver disease, cancers, acute alcohol poisoning, and fetal alcohol syndrome. Alcoholism and other health related issues like smoking are known to be influenced by one's social environment~\cite{gal}. With increase in usage of online social media as a preferred medium of communication, it has become a diagnostic tool to identify human nature. According to Pew Research Center, as of January 2014, 74\% of online adults use social networking sites; the number is more than 80\% for individuals under the age of 50. Also from the reports published by the Centers for Disease Control and Prevention (CDC)\footnote{http://1.usa.gov/23PMj4F}, we found the prevalence of heavy drinkers/smokers in the said age group. This suggests that social media is a viable platform to study the alcoholic users and the interaction (exchange of messages, posts etc.) in these social media has opened up a research corridor for observing and understanding individuals' psychological states and their social environment. It is very important to identify how these characteristics vary dynamically for different human behaviors. It will also be quite informative to examine how different characteristics vary demographically (sex, age, region etc.), for different time frames like days of week (weekdays vs weekends), monthly (start of the month vs end of the month) or hourly (morning vs work hours vs evening vs late night). Demographic patterns can be different for psychogenic people, predominant drunkers and others scenarios than the normal people. For example, we can identify predominant drunk peoples' suicidal tendencies or change in behavior in near future by tracking social media so that we can control situations accordingly. Thus we can use social media as an important medical diagnostic system and develop a predominant drunker identification model. 
	
	In this paper, we investigate how social media language usage and interactions can be used to characterize and understand the drunk texters. Subsequently, we leverage on the behavioral, social, psychological and linguistic aspects of the Twitter users to propose a classification framework to automatically identify the drunk texters. The automatic identification of drunk texters is important because these users can then be targeted by the communities that are missioned to cure alcohol abuse and help the alcoholics to quit addiction. Also as these users tend to abuse in social media under the influence of alcohol, our automatic identification framework can be used to enrich the process of filtering abusive contents from the media.
	
	\section{Related Work}
	There have been several works on health and social media. Joshi et al.~\cite{joshi2015computational} propose a computational framework for identifying drunk tweets from non-drunk tweets. Tamersoy et al.~\cite{tamersoy2015characterizing} study the abstinence from smoking and drinking. They use linguistic features of the content shared by the users as well as the network structure of their social interactions to distinguish between the short-term and long-term abstinence. Murnane and Counts\cite{murnane2014unraveling} examine the cessation process of smoking. Many of the past research works focus on finding relationship between alcohol abusers with human aggression~\cite{bushman1990effects}, crime~\cite{carpenter2007heavy}, suicide~\cite{merrill1992alcohol}. 
	
	Strapparava and Mihalcea~\cite{strapparava2017computational} perform a computational analysis of the language of drug users when talking about their drug experiences. Cameron et al.~\cite{cameron2013predose} develop a web platform (PREDOSE), focusing on epidemiological study of prescription and related drug abuse practices using social media (e.g., online forums). Paul and Dredze~\cite{paul2012experimenting,paul2013drug} have developed multidimensional latent text model to capture orthogonal factors that correspond to drug type, delivery method (smoking, injection, etc.), and aspect (chemistry, culture, effects, health, usage). Coyle et al.~\cite{coyle2012quantitative} classify and characterize different kinds of drug use experiences, using a random-forest classifier over 1000 reports of 10 drugs from the drug information website Erowid.org (manually identified subsets of words differentiated by drugs). 
	
	On the other hand, there exist several works that try to establish the role of online social media in alcoholic's life - how it influence alcohol use of the adult users~\cite{cook2013online}, how people use the social network to display their drunk behaviors~\cite{beullens2013display}. West et al.~\cite{west2012} examine the extent to which individuals tweet about the problem of drinking, and to identify if such tweets correspond with time periods when the problem of drinking was likely to occur.
	
	Some researches focused on extracting various sociological aspects from online social media. Coppersmith et al.~\cite{coppersmith2015adhd} analyze broad range of mental health conditions in Twitter texts by identifying self-reported statements of diagnosis. Schwartz et al.~\cite{schwartz2013personality} predict latent personal attributes including user demographics, online personality, emotions and sentiments from texts published on Twitter. Volkova et al.~\cite{volkova2015inferring} explore emotion, sentiment and other personality types.
	
	\section{Dataset Preparation}
	Our first step was to identify Twitter users who are drunk texters. To achieve this goal, we used the manually labelled tweet dataset mentioned in~\cite{hossain2016inferring}. We then separately crawled the timeline of the posters of these tweets. We then filtered out the tweets of these users based on keyword\footnote{Initial seed keywords are collected from \cite{west2012} like - `drunk', `tipsy', `intoxicated', `buzzed' etc. Later we increased the datasets using similar keywords from wordnet like `booze', `juiced' etc. and make the final wordset of length 61.} and then got the tweets manually labelled as drunk-texts or not by 3 of the authors. We considered only those tweets which are tagged drunk text unanimously by all of them. After this manual labeling, we consider those users who have posted at least 5 drunk tweets. In total, we had 278 drunk texters. We then prepared the dataset corresponding to the non-drunk texters\footnote{Normal users are defined as the user who never posted any `drunk' related tweets i.e. none of the tweet contain any word from the previous wordset of length 61.}. We use Twitter 1\% random sample from the month of January, 2014 to obtain a set of users who didn't have any tweets containing any of the keywords related to alcohol consumption. We chose 278 such non-drunk texters from this set in order to keep both the sets comparable. Following are the example tweets which depict that the user is a drunk-texter.
	\begin{itemize}
		\item I know its Saturday but I'm trying to get roofied drunk
		\item Gotta say, my spelling's been pretty on-point considering how drunk I've been tonight
		\item Alcohol and weed are like the mom and dad I always wanted
	\end{itemize}

	\section{Behavioral, Psychological and linguistic aspects of the drunk texters}
	In this section, we focus on the comparative study of the drunk texters and non-drunk texters based on their behavioral, psychological and linguistic aspects. Our empirical study is based on the content extracted from the tweets of the drunk and non-drunk texters. Each of the analysis has been done separately for the tweets posted on weekends and weekdays to differentiate between the lifestyles of the users over the weekdays and in the weekends.
	
	\subsection{Health and food}
\begin{figure}[!tbh]
		%\vspace{-2mm}
		\centering
		\includegraphics[scale=0.2]{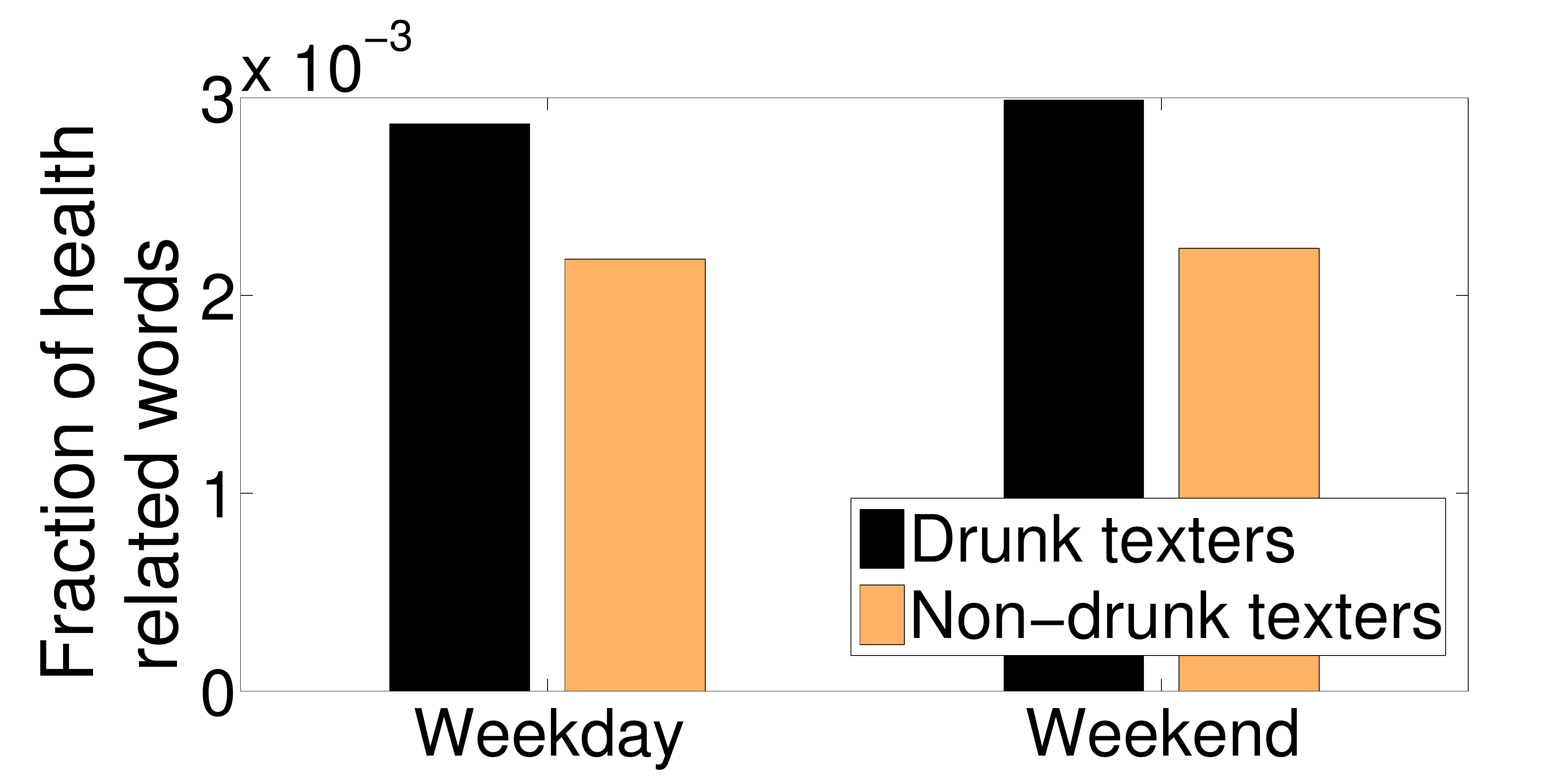}
		%\captionof{figure}{Only Picture}
		%\vspace{-24mm}
		\caption{Health}
		\label{fig:health}
		%\vspace{-3mm}
	\end{figure}

\begin{figure}[!tbh]
		%\vspace{-2mm}
		\centering
		\includegraphics[scale=0.2]{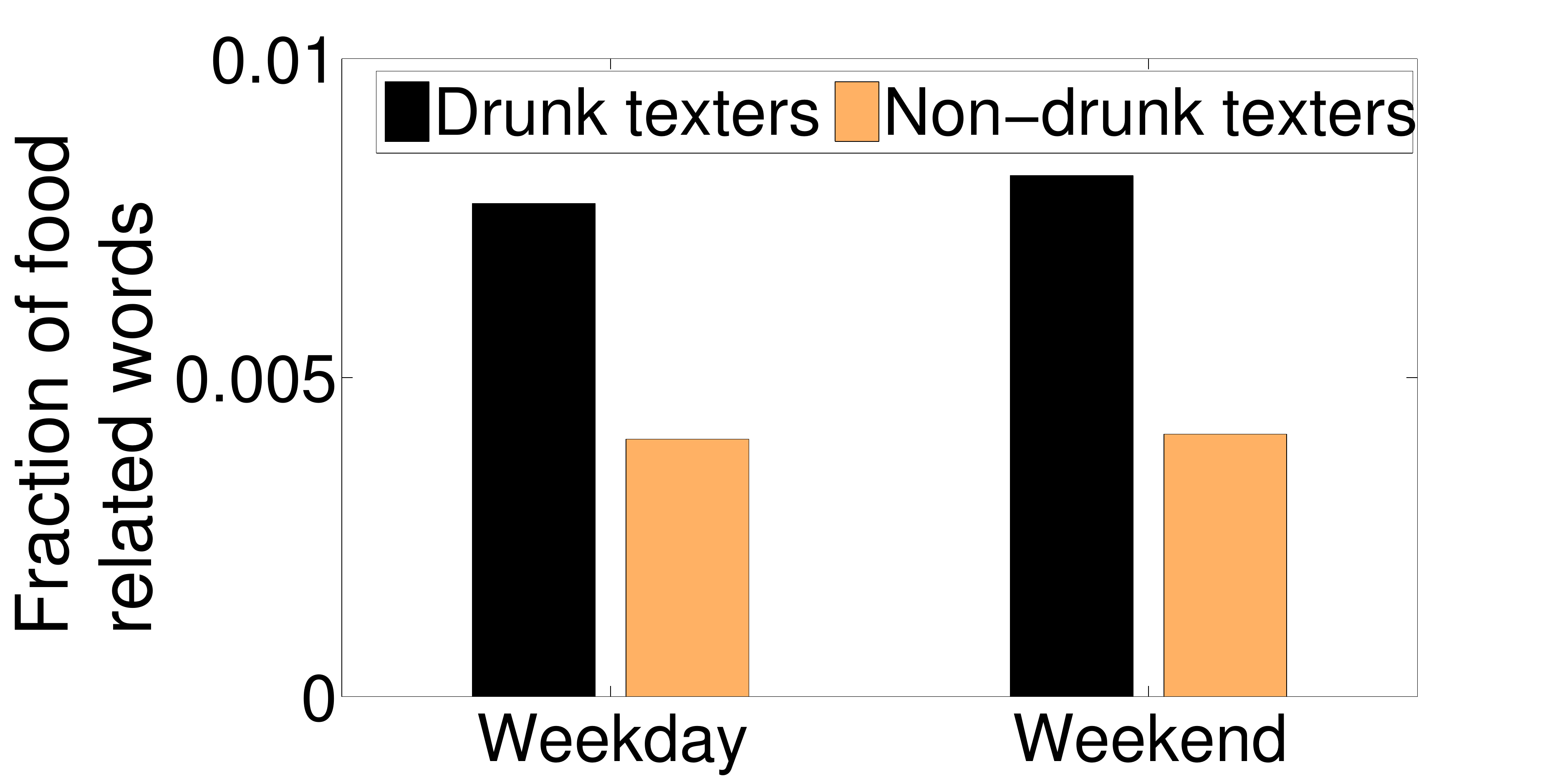}
		%\captionof{figure}{Only Picture}
		%\vspace{-24mm}
		\caption{Food}
		\label{fig:food}
		\vspace{-3mm}
	\end{figure}

	\iffalse People post a lot of messages on Twitter related to day-to-day activities,\fi Since health is one of the crucial aspects of well-being, people often share information related to health and food over social media. We empirically find if drunk texters and non-drunk texters have some contrasting contents related to health and food. Consumption of alcohol has adverse impacts on health. It could be long-term (impact on health over a period of time) or short-term (hangover from last night or throwing up)\footnote{http://1.usa.gov/1d7aWk2}; so drunk texters might share their experiences on Twitter. To obtain the behavior of drunk texters and non-drunk texters in regard to health and food content sharing, we compiled a list of most frequently used health and food related keywords\footnote{\label{note1}http://bit.ly/200kea3} on social media; further we computed the fraction of health and food related keywords for both the set of users. Figure \ref{fig:health} and \ref{fig:food} show that drunk texters, in general, use more of health and food related keywords in their tweets as compared to non-drunk texters.
	
	\subsection{Stress}
	People tend to drink in response to stress, accordingly exposure to the tension-producing situations lead to increased drinking \cite{citer7}; so there is a high chance that drunk texters while posting the tweets will communicate their stress. In general stress levels are rising severely, a survey by American Psychological Association portrays a picture of high stress and ineffective coping mechanisms that appear to be ingrained in our culture\footnote{http://bit.ly/1cz4n99}. People might share the stressful situations they have been in, so non-drunk texters also have a decent chance of posting tweets expressing stress and anxiety.
	
\begin{figure*}[h]
		\centering
		\includegraphics[scale=0.32]{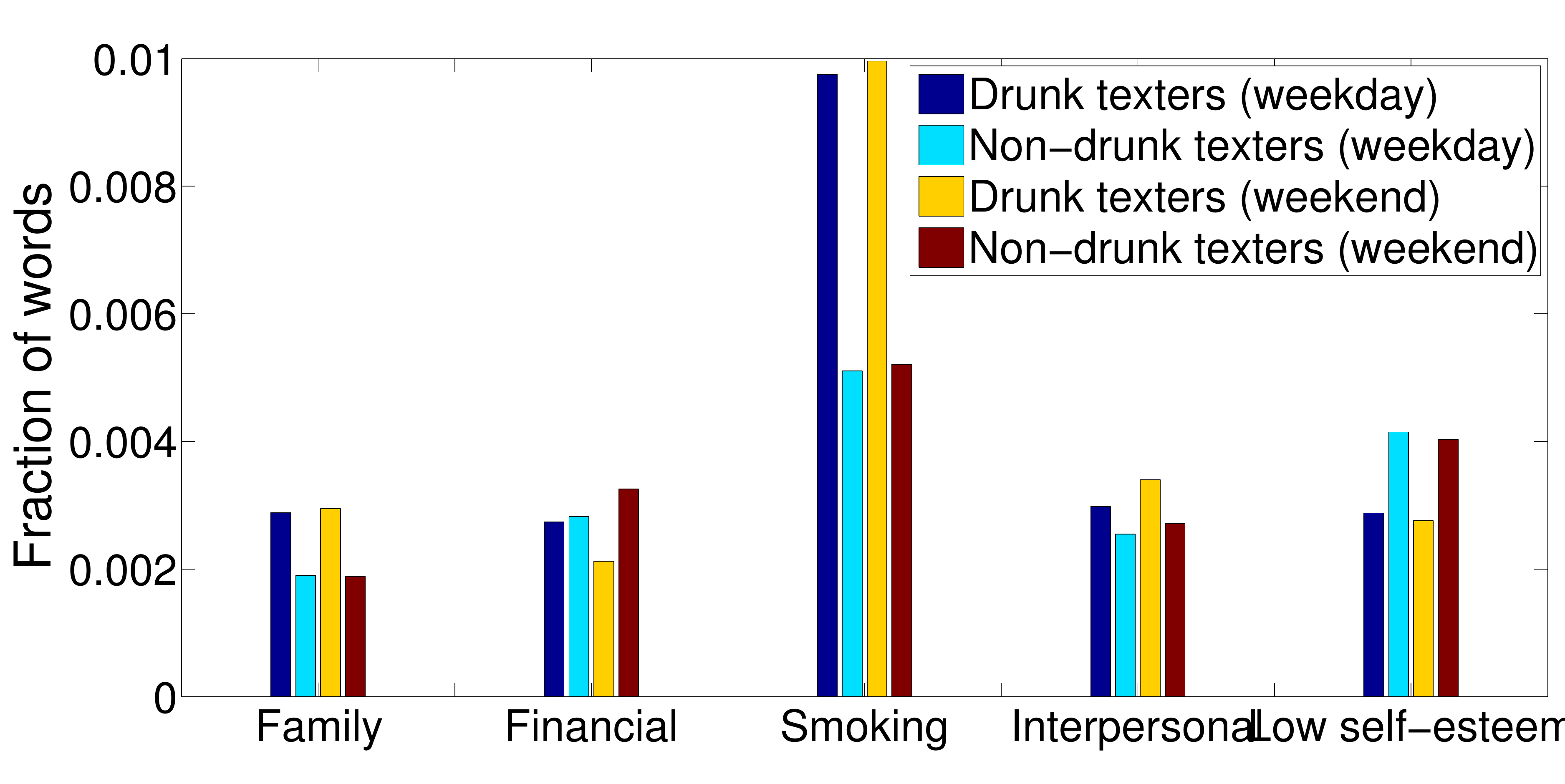}
		
		\caption{ Different sources of stress (y-axis values are scaled up by 10 times in case of financial and low self-esteem stress for better visualization.)}
		\label{fig:stress}
	\end{figure*}

	The major sources of stress are listed as follows \cite{citer8} 
	\begin{itemize}
		\item Low Self-esteem
		\item Inter-personal conflicts
		\item Smoking
		\item Financial difficulties
		\item Family problems
	\end{itemize}
	To empirically find the stress related behavior of the drunk and the non-drunk texters we gather a list of stress related keywords\footnotemark[\getrefnumber{note1}] corresponding to each of the source of the stress mentioned above. Further we compute the fraction of stress related keywords for both the drunk and non-drunk texters. Figure~\ref{fig:stress} shows the contrasting behavior between them and illustrates that in general non-drunk texters seem to experience more stress arising out of financial problems and low self-esteem whereas drunk texters experience more stress due to inter-personal conflicts, smoking and family problems.
	
	% \begin{figure}[!ht]
	% \vspace{-.3cm}
	% \centering
	% \subfigure[low self esteem]{\label{fig:low self esteem}{\includegraphics[height=0.15\textwidth]{low_self_esteem}}}
	% \hspace{0em}
	% \subfigure[interpersonal conflicts]{\label{fig:interpersonal}{\includegraphics[height=0.15\textwidth]{interpersonal}}}
	% \hspace{0em}
	% \subfigure[smoking]{\label{fig:smoking}{\includegraphics[height=0.15\textwidth]{smoking}}}
	% \hspace{0em}
	% \subfigure[financial]{\label{fig:financial}{\includegraphics[height=0.15\textwidth]{financial}}}
	% \hspace{0em}
	% \subfigure[family]{\label{fig:family}{\includegraphics[height=0.15\textwidth]{family}}}
	% \vspace{-.3cm}
	% \caption{Stress}
	% \label{fig:stress}
	% \vspace{-.5cm}
	% \end{figure}
	
	% \begin{figure}[h]
	%     \centering
	% \includegraphics*[scale = 0.13,angle=0]{money.eps}
	% \caption{\label{fig:money} Money related words}
	% \end{figure}

	\subsection{Swearing and abusing}
\begin{figure}[h]
		\centering
		\includegraphics[scale=0.2]{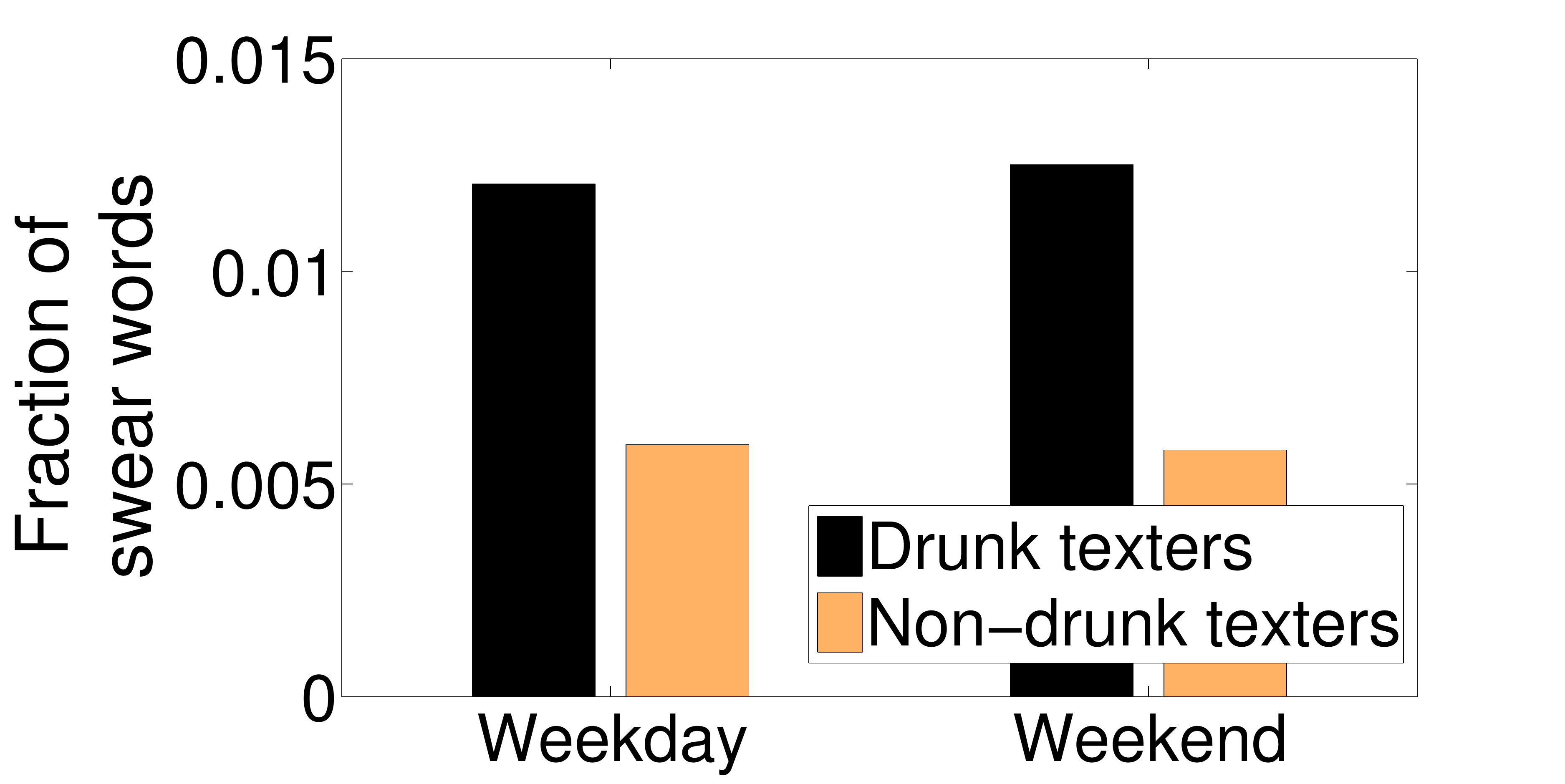}
		\caption{\label{fig:swear} Swear words}
        \vspace{-4mm}
	\end{figure}

	Alcohol consumption is closely related to violent behavior\cite{Obot2000169, Greenfeld}. Swearing being a verbal form of aggression can serve as an indicator of aggressive behavior. We speculate that drunk texters in general are more probable to use swear words in their tweets because of relatively higher violent behavior. To investigate whether this trend is also observed on Twitter we compiled a list of swear related keywords\footnotemark[\getrefnumber{note1}] used most frequently on social media and then compute the fraction of such keywords for both the drunk and the non-drunk texters. Figure~\ref{fig:swear} supports our speculation that drunk texters use a larger proportion of swear words in their tweets compared to non-drunk texters.
	
	% \begin{figure}[!ht]
	% \vspace{-.5cm}
	% \centering
	% \includegraphics[height=0.15\textwidth]{swear}
	% \vspace{-.3cm}
	% \caption{Swearing}
	% \label{fig:swear}
	% \vspace{-.5cm}
	% \end{figure}
	
	%\begin{figure}[!ht]
	%\begin{minipage}[b]{0.20\textwidth}
	%  \centering
	%  \includegraphics[width=\linewidth]{money_n.eps}
	%  \vspace{-.3cm}
	%  \caption{\small Money}
	%  \label{fig:money}
	%  \vspace{-3mm}
	%\end{minipage}
	%\hspace{0.2cm}
	%\begin{minipage}[b]{0.20\textwidth}
	%  \centering
	%  \includegraphics[width=\linewidth]{swear_n.eps}
	%  \vspace{-.3cm}
	%  \caption{\small Swearing}
	%  \label{fig:swear}
	%  \vspace{-3mm}
	%\end{minipage}
	%\end{figure}
	% 
	\subsection{Money}
\begin{figure}[!ht]
		\centering
		\includegraphics[scale=0.2]{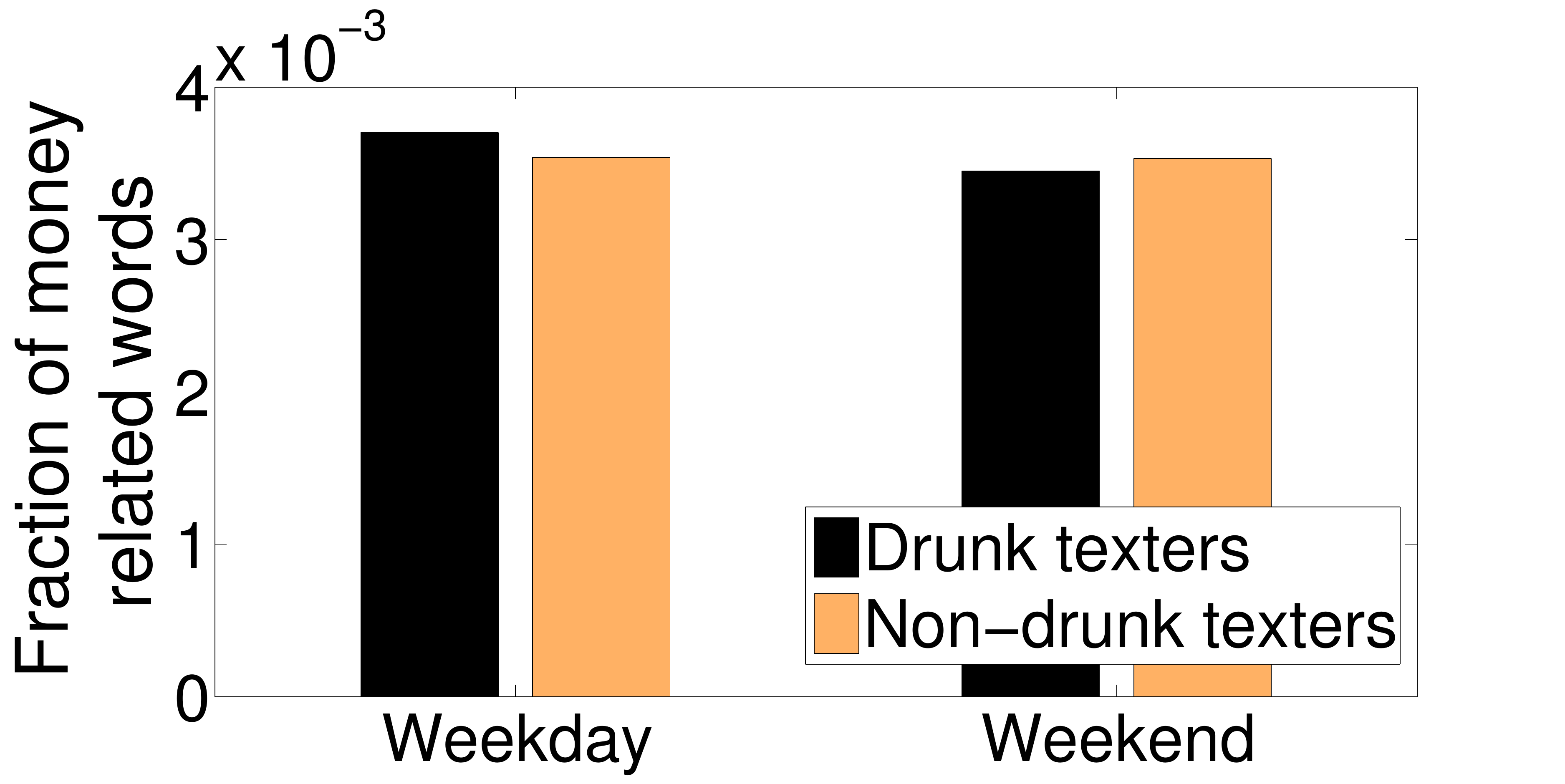}
        \vspace{-2mm}
		\caption{Money}
		\label{fig:money}
        \vspace{-6mm}
	\end{figure}

	Spending money and drinking alcohol are positively correlated \cite{citer4}. Drunk texters might post about their spending on drinks which might be a considerable share of their income. For the analysis, we compiled a list of money related keywords\footnotemark[\getrefnumber{note1}] used most frequently on social media and then computed the fraction of money related keywords for both the alcoholic and the non-drunk texters. Figure \ref{fig:money} shows that drunk texters are more likely to use money related words during the weekdays compared to the weekends in their tweets.
	
	\subsection{Sentiment analysis}
	% \noindent{\bf Opinions:} Twitter users generate a lot of raw data posting real-time messages on a variety of topics in daily life. These messages can be opinions or some factual information. We envisage to obtain some differences among the alcoholic and non-drunk texters on the kind of messages they post. To take this into consideration we used opinion mining techniques \cite{citer9} on the tweets posted by the alcoholic and non-drunk texters. Figure \ref{fig:opinion} shows the behavior of alcoholic and non-drunk texters and illustrates that in general non-drunk texters have higher opinion score in their tweets as compared to drunk texters.
	
\begin{figure}[h]
		\centering
		\includegraphics*[scale=0.2]{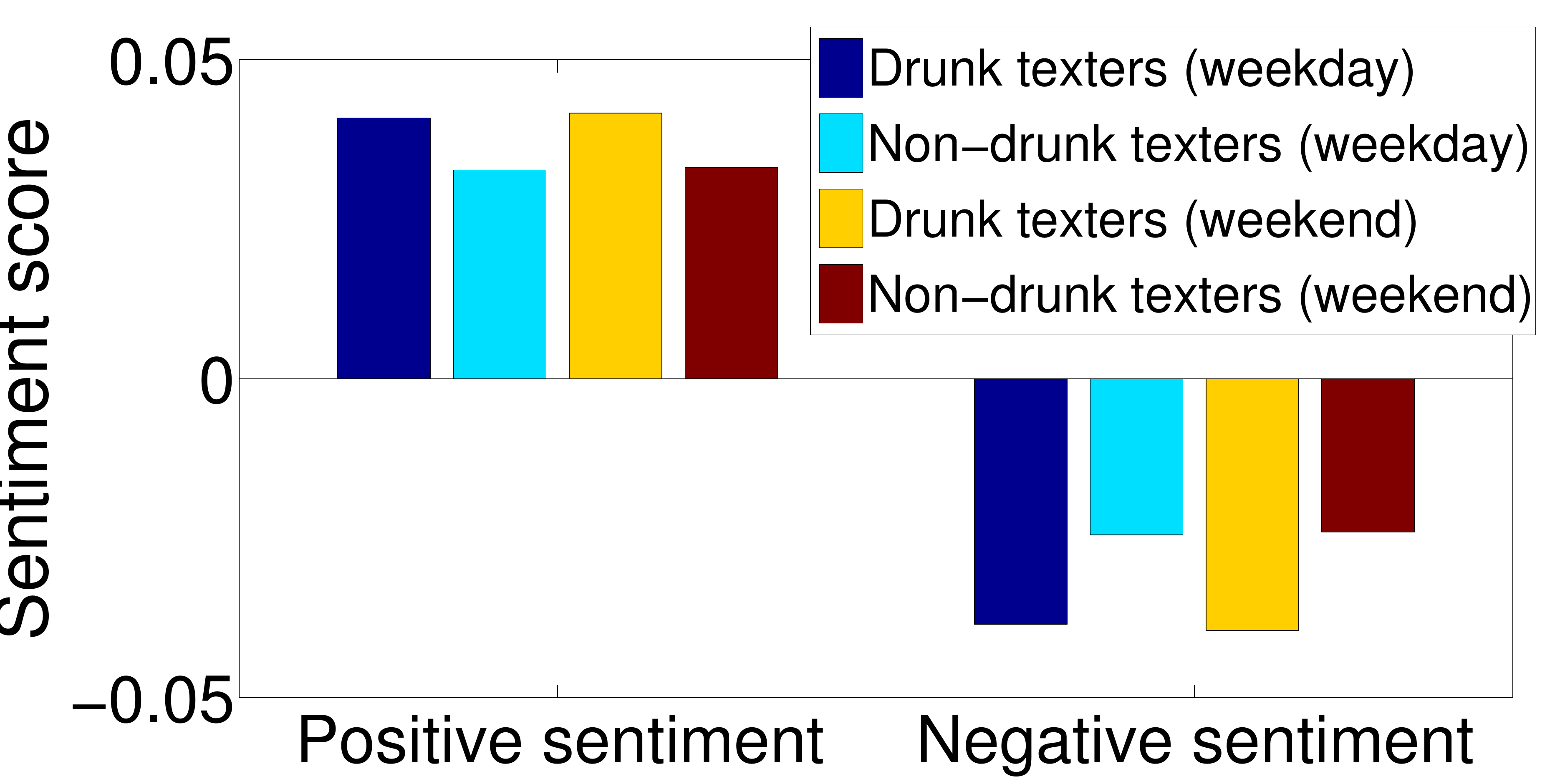}
		\caption{\label{fig:pos_neg} Sentiment scores}
	\end{figure}

	% \begin{figure}[!hb]
	% \centering
	% \includegraphics[height=0.15\textwidth]{opinion}
	% \caption{opinion}
	% \label{fig:opinion}
	% \end{figure}
	Sentiments of a user greatly depend on the state of the user. We believe that a user's tweets shall largely depend on the state in which the user is tweeting. People tend to speak differently when he/she is in a drunken state compared to when in a normal state. The same clause should be applicable while the user is tweeting. We have used sentiment lexicon~\footnote{\url{https://www.cs.uic.edu/~liub/FBS/sentiment-analysis.html\#lexicon}} for the sentiment analysis. Figure~\ref{fig:pos_neg} shows the behavior of the drunk and the non-drunk texters and illustrates that in general drunk texters have higher sentiment score in their tweets as compared to non-drunk texters.
	
	% \begin{figure}[!ht]
	% \vspace{-.3cm}
	% \centering
	% \subfigure[opinion]{\label{fig:opinion}{\includegraphics[height=0.15\textwidth]{opinion}}}
	% \hspace{0em}
	% \subfigure[positive sentiment]{\label{fig:positive}{\includegraphics[height=0.15\textwidth]{positive}}}
	% \hspace{0em}
	% \subfigure[negative sentiment]{\label{fig:negative}{\includegraphics[height=0.15\textwidth]{negative}}}
	% \vspace{-.3cm}
	% \caption{Opinion and sentiment analysis}
	% \label{fig:sentiment}
	% \vspace{-0.5cm}
	% \end{figure}
	
	%\begin{figure}[!ht]
	%\centering
	%\subfigure[opinion]{\label{fig:opinion}{\includegraphics[height=0.11\textwidth]{opinion_n.eps}}}
	%\hspace{0em}
	%\subfigure[sentiment]{\label{fig:positive}{\includegraphics[height=0.11\textwidth]{pos_neg_n.eps}}}
	%%\hspace{0em}
	%%\subfigure[negative sentiment]{\label{fig:negative}{\includegraphics[height=0.15\textwidth]{negative}}}
	%\vspace{-.3cm}
	%\caption{\small Opinion and sentiment analysis}
	%\label{fig:sentiment}
	%\end{figure}
	%\vspace{-4mm}
	
	\subsection{Psychological and linguistic states}
	Theories on drinking and aggression postulate that alcohol contributes indirectly to increased aggression by causing cognitive, emotional, and psychological changes that may reduce self-awareness or result in inaccurate assessment of risks~\cite{citer3}. The function and emotion words people use provide important psychological cues to their thought processes, emotional states, intentions, and motivations. To capture user's social and psychological states we used Linguistic Inquiry and Word Count (LIWC) framework~\cite{citer5}. Some of the interesting observations are presented in Table~\ref{tab:liwc}. It is evident from the table that drunk texters express more anxiety, anger, sadness and also show more sexual aggression by using more sexual words in their tweets than the non-drunk texters. Also the drunk texters tweet more about leisure activities and are less religious.
	
	\begin{table}[h]
		\caption{Psycholinguistic analysis for drunk and non-drunk texters. $\alpha$, $\beta$, $\gamma$, $\delta$ are avg. LIWC scores for drunk texters on weekday, non-drunk texters on weekday and drunk texters on weekend, non-drunk texters on weekend respectively.}
		\vspace{2mm}
		\label{tab:liwc}
		% \centering
		%  \resizebox{8.5cm}{!}{
		\begin{tabular}{|p{3cm}|c|c|c|c|}
			\hline\
			LIWC category & $\alpha$ & $\beta$ & $\gamma$ & $\delta$\\
			\hline
			% \multicolumn{5}{|c|}{Linguistic Categories} \\ \hline
			% Dictionary words & 45.72 & 52.15 & 45.99 & 52.12\\
			% Total function words & 25.56 & 29.88 & 25.76 & 29.81\\
			% Total pronouns & 8.01 & 9.21 & 8.03 & 9.13\\
			% Personal pronouns & 5.73 & 6.48 & 5.76 & 6.42\\
			% Articles & 2.83 & 3.10 & 2.82 & 3.11\\
			% Common verbs & 10.18 & 9.97 & 10.37 & 9.93\\
			% Prepositions & 5.67 & 6.86 & 5.73 & 6.84\\
			% Conjunctions & 2.31 & 2.73 & 2.31 & 2.75\\\hline
			% \multicolumn{5}{|c|}{Psychological Categories} \\ \hline
			Social processes & 8.69 & 6.88 & 8.86 & 6.78\\
			Family & 0.4 & 0.27 & 0.48 & 0.29\\
			Friends & 0.28 & 0.17 & 0.31 & 0.17\\
			Anxiety & 0.33 & 0.22 & 0.30 & 0.22\\
			Anger & 1.55 & 0.79 & 1.62 & 0.78\\
			Sadness & 0.50 & 0.34 & 0.52 & 0.33\\
			Body & 1.22 & 0.68 & 1.24 & 0.68\\
			Sexual & 1.10 & 0.61 & 1.19 & 0.57\\
			Ingestion & 0.79 & 0.36 & 0.83 & 0.35\\
			% Work & 1.45 & 1.29 & 1.11 & 1.17\\
			% Achievement & 0.95 & 1.34 & 0.93 & 1.33\\
			Leisure & 1.83 & 1.42 & 2.14 & 1.56\\
			Religious & 0.37 & 0.41 &0.36 & 0.42 \\
			\hline
		\end{tabular}
		% }
	\end{table}

	\section{Classification Framework}
	From discussions in the earlier section, it is evident that there exists differences between drunk and non-drunk texters in various behavioral, psychological and linguistic aspects. We use these discriminative aspects as features in our classification framework to classify a user into a drunk texter or not. We use 10-fold cross-validation technique of various classifiers like Support Vector Machines (SVM), Logistic Regression (LR), Random Forest (RF), Bagging, Decision Tree (DT-J48), Na{i}ve Bayes, Ada Boost for checking robustness of our method. All the classifiers perform very well. Table~\ref{tab:methods} shows that the evaluation results for weekday and weekend data with various classification techniques in terms of accuracy, precision, recall, F1-Score, ROC Area. SVM classifier performs the best as we obtain 96.78\% (weekday), 96.14\% (weekend) accuracy with avg. precision - 0.968 (weekday), 0.963 (weekend) and recall of 0.968 (weekday) and 0.961 (weekend). It also gives better area under the ROC curve. We also compared the drunk texters set with a random sample of users and we achieve a similar very high accuracy with high precision and recall which establishes the fact that the features we use are robust and strong discriminators of drunk-texting.
	\iffalse{}	
	\begin{table}[h]
		\small
		\centering
		\caption{Evaluation results for various classifiers - Support Vector Machines (SVM), Logistic Regression (LR), Random Forest (RF).}
		\label{tab:methods}
		% \centering
		%  \resizebox{7.5cm}{!}{
		\begin{tabular}{ |c|p{1cm}|p{1cm}|p{1cm}| }
			\hline
			Metrics & SVM &LR & RF \\ \hline
			Accuracy & 96.62\% & 96.46\% & 95\% \\\hline
			Precision & 0.967 & 0.965 &0.95  \\\hline
			Recall & 0.966 & 0.965 & 0.95 \\\hline
			F-Score & 0.966 & 0.965& 0.95 \\\hline
			ROC Area & 0.968 & 0.987 & 0.989 \\\hline
			
			% Classifier & Accur-acy &Preci-sion & Recall & F-Score & ROC Area \\ \hline
			%   Support Vector Machine & 96.62\%& 0.967   &  0.966   &  0.966   &   0.968 \\\hline
			%   Logistic Regression & 96.46\% &0.965   & 0.965   &  0.965    &  0.987 \\\hline
			%   Random Forest & 95\% & 0.951   &  0.95  &    0.95    &   0.989 \\\hline
		\end{tabular}
		%  }
	\end{table}
	\fi	
	
	\begin{table*}[!htb]
		\centering
		\caption{Evaluation results for various classifiers - Support Vector Machines (SVM), Logistic Regression (LR), Random Forest (RF), Bagging, Decision Tree (DT), Naive Bayes (NB), Ada Boost in terms of Accuracy (Acc.), Precision (P), Recall (R), F1-Score (F1) and Area under Receiver operating characteristic (ROC) curve for weekday and weekend data.}
		\label{tab:methods}
		\begin{small}
			\begin{adjustbox}{width=0.985\linewidth,center}
				\begin{tabular}{|c|c|c|c|c|c||c|c|c|c|c|}
					\hline
					& \multicolumn{5}{c|}{Weekday} & \multicolumn{5}{|c|}{Weekend} \\ \hline
					Classifiers		 & Acc. (\%)    & P    & R    & F1    & ROC   & Acc. (\%)     & P   & R   & F1   & ROC  \\ \hline
					\textbf{SVM}			 &\textbf{96.78} & \textbf{0.968} & \textbf{0.968} & \textbf{0.968} & \textbf{0.991}  & \textbf{96.14}&\textbf{0.963} &\textbf{0.961} &\textbf{0.962} & \textbf{0.994}  \\ \hline
					LR				 &96.62 & 0.967 & 0.966 & 0.966 & 0.986  & 95.17&0.952 &0.952 &0.952 & 0.991  \\ \hline
					RF				 &95.81 & 0.958 & 0.958 & 0.958 & 0.987  & 94.85&0.949 &0.948 &0.948 & 0.989  \\ \hline
					Bagging			 &94.04 & 0.941 & 0.94 & 0.94 & 0.984  & 95.01&0.95 &0.95&0.95 & 0.981  \\ \hline
					DT(J48)			 &94.36 & 0.944 & 0.945 & 0.945 & 0.948  & 93.88&0.939 &0.939 &0.939 & 0.918  \\ \hline
					NB				 &91.46 & 0.92 & 0.915 & 0.917 & 0.971  & 90.18&0.91 &0.90 &0.905 & 0.967  \\ \hline
					Ada Boost		 &94.68 & 0.947 & 0.947 & 0.947 & 0.988  & 95.49&0.955 &0.955 &0.955 & 0.988   \\ \hline
					
				\end{tabular}
			\end{adjustbox}
		\end{small}
	\end{table*}
	
	In order to determine the discriminative power of each feature, we compute the chi-square ($\chi^2$) value and the information gain. Table~\ref{tab:featurerank} shows the rank order of all features based on the $\chi^2$ value. The ranks of the features are very similar when ranked by information gain (Kullback-Leibler divergence). The most prominent discriminative features are various linguistic as well psychological features obtained from LIWC.
	
	\begin{table}[!htb]
		% \tiny
		% \small
		\centering
		\
		\caption{\small Top 25 predictive features and their discriminative power}
		\vspace{2mm}
		\label{tab:featurerank}
		% \resizebox{8.5cm}{!}{
		\begin{tabular}{ |p{1.8cm}|p{1cm}|c| }
			\hline
			$\chi^2$ Value & Rank  & Feature  \\ \hline
			494.8247  &  1 & Dictionary words \\
468.8026   & 2 & Function words \\
407.5107    & 3 & Relativity \\
395.2954   & 4 & Adverbs \\
391.3315   & 5 & Time \\
381.0396  & 6 & Ingestion \\
373.1287   & 7 & Space \\
363.9906    & 8 & Inclusive words \\
353.0703    & 9 & Cognitive processes \\
352.5059   & 10 & Auxiliary verbs \\
349.9656   &  11 &  Preposition \\
342.1795   & 12 & Common verbs \\
328.7726   & 13 & Smoking related words\\
322.1683   & 14 &  Biological \\
318.6225   & 15 & Conjunctions \\
316.1245   & 16 & Present tense \\
314.3415   & 17 & Pronouns \\
312.5311   & 18 & Past tense \\
311.6849   & 19 & $1^{st}$ person singular \\
304.3776    & 20 & Home related words \\
294.9148    & 21 & Quantifiers \\
292.5086    & 22 & Impersonal pronouns \\
292.2972    & 23 & Motion related words \\
289.969     & 24 & Food related words \\
281.1014    & 25 &  Certainty \\ \hline
		\end{tabular}
		% }
	\end{table}
	
\vspace{-5mm}
\section{Discussions}
	\subsection{Bot Detection}
\begin{figure}[!ht]
		\centering
		\includegraphics[width=0.9\columnwidth,height=5cm]{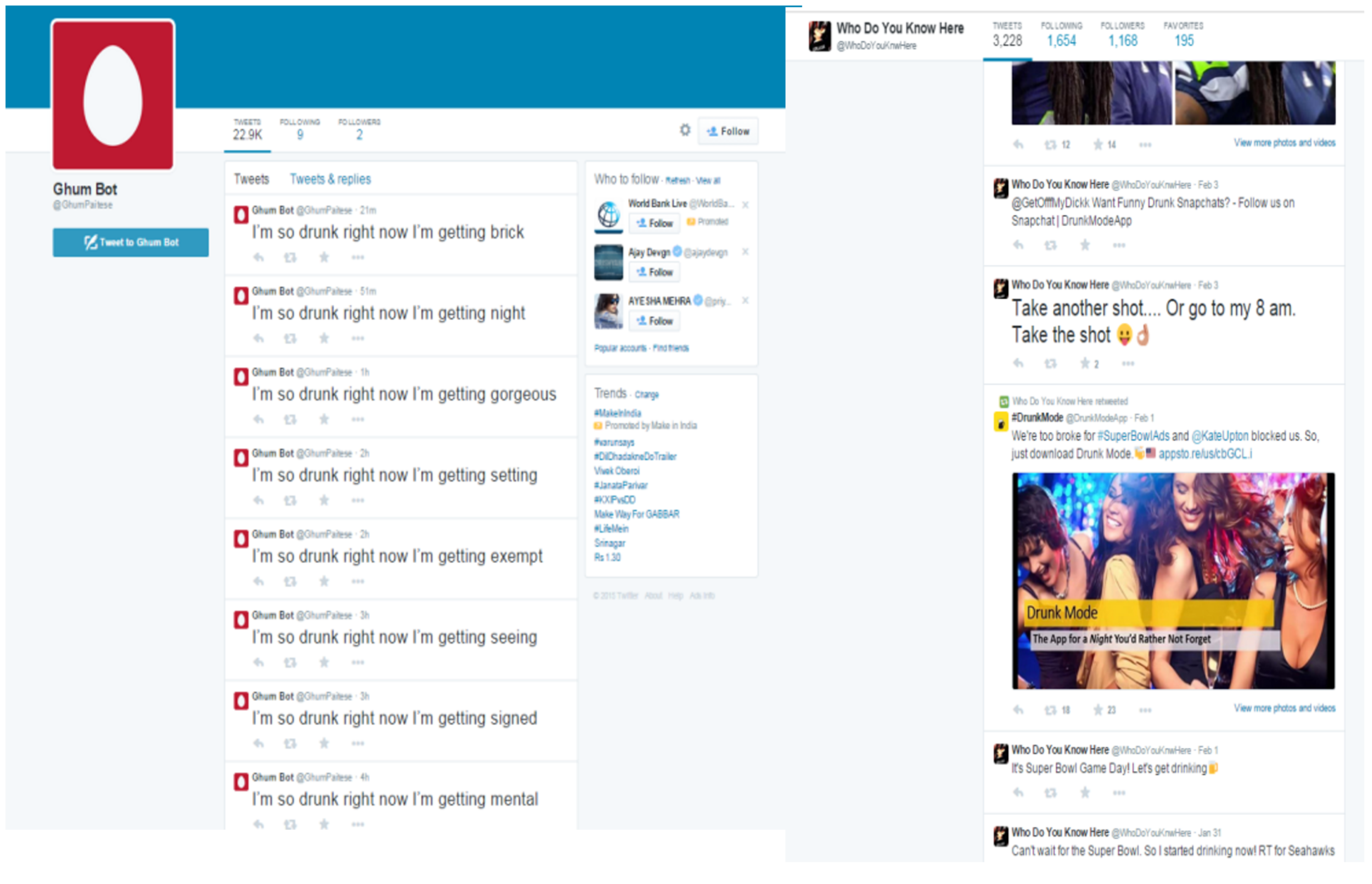}
		\caption{Drunk texting Bots}
		\label{fig:bot}
        \vspace{-4mm}
	\end{figure}

	We have identified bots having more than 99\% drunk related tweets, for example - `GhumPaitase', `WhoDoYouKnwHere', `UrDrunkTweets' etc. Our system were also able to detect bots as shown in Fig. \ref{fig:bot}.
	
	\subsection{Temporal Tweeting behavior and community detection}
\begin{figure*}[!htb]
		\centering
		\includegraphics[scale=0.1]{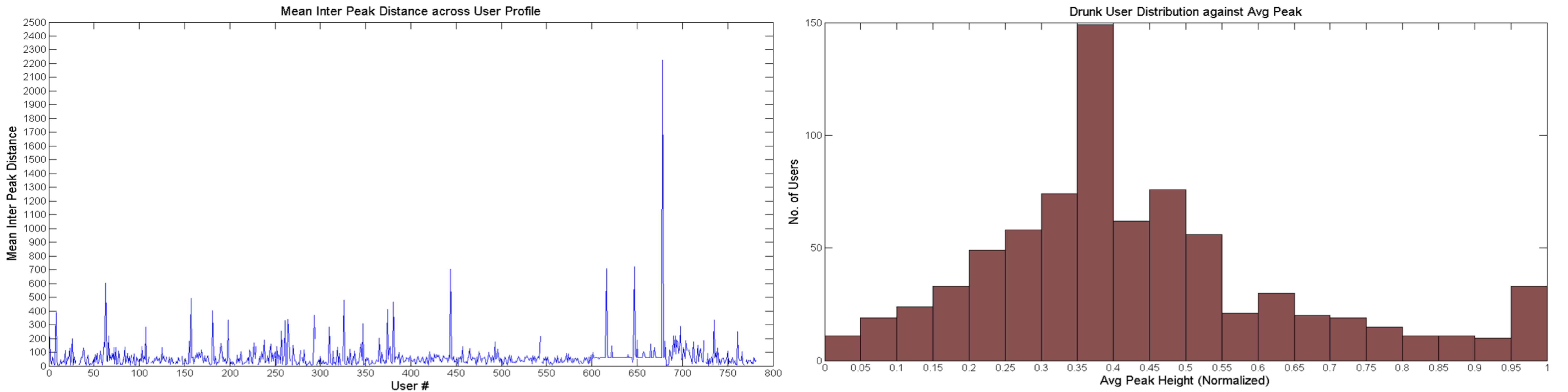}
		\caption{Peak Analysis}
		\label{fig:peak}
        \vspace{-3mm}
	\end{figure*}

	We further try to understand the temporal tweeting pattern of the users\footnote{To capture the temporal tweeting characteristics more efficiently, we increase the drunk texters's dataset to $\sim$800 users}. For this task, we identify some additional keywords, based on their co-occurrences with drunk words (61 length wordset) and we assign each tweet a `drunk' score based on these words and then analyze the peaks in the profile as shown in Fig. \ref{fig:peak}. We observe that - (i) average peak height of tweets of drunk texters follow normal distribution, (ii) most of the drunk texters having inter-peak distance less than 100 tweets.
    
\textbf{Existence of communities} We also study the existence of communities among the drunk texters. We identify 2 different types of communities : \\
	1. \textbf{Interest Based Communities}: \\
   First, we investigate whether there exist interest-driven communities. For this task, for each user, we construct a vector of the features - (a) no. of peaks, (b) average peak height (c) std. error (peak height) (d) max peak height (e) mean peak interval and (f) std. error (peak interval). Users are the nodes in the graph and an edge between two users are formed if the cosine similarity of the feature vectors of the user-pair crosses a certain threshold (0.2). We then apply Louvain Algorithm to detect communities. Three different types of communities are formed of length - 276, 193 and 312.\\
	2. \textbf{Bond Based Communities}: \\
  We also observe that these users have common friends and followers and the distribution shows a power-law behavior. Hence, we try to observe if there are social communities formed among these drunk texters. We construct two kind of communities - based on common friends and common followers. For common friends-based communities, we obtain a total of 179 communities and for common followers-based communities, 283 communities are formed which suggest that there are large number of small-sized communities existing.
	 
\section{Conclusion}
	In this paper, we investigate various psycholinguistic aspects of the drunk texters.
   We then use these characteristic properties as features for a classification model that tries to classify whether a user is drunk texter or not. To the best of our knowledge, this is the first study which tries to use the psycholinguistic aspects of social media interactions to identify drunk texters. Our proposed classification framework achieves an accuracy of 96.78\% (weekday), 96.14\% (weekend) with very high precision and recall. This high accuracy suggest that it can be used as an alternate approach for identifying keyword-based classification of drunk texters which requires a lot of manual intervention to obtain accurate results. We observed that linguistic features (LIWC) are the most discriminative features compared to others. 
	One immediate future research is to identify various steps of how social media influence a non drunk person to become predominant drunkers and by detecting change in characteristics in various demographic dimensions how can we increase social awareness to decrease social influences. One direction is to explore different feature behaviors - like how opinion dynamics \cite{mullick2017generic} change or correlation with other different addictions for predominant drunkers compared to non-drunkers. Another idea is to detect various subsets of drunkers - occasional, situational or regular and respective change in personal life and different associated health hazards. 
	
	% trigger a \newpage just before the given reference
	% number - used to balance the columns on the last page
	% adjust value as needed - may need to be readjusted if
	% the document is modified later
	%\IEEEtriggeratref{8}
	% The "triggered" command can be changed if desired:
	%\IEEEtriggercmd{\enlargethispage{-5in}}
	
	% references section
	
	% can use a bibliography generated by BibTeX as a .bbl file
	% BibTeX documentation can be easily obtained at:
	% http://mirror.ctan.org/biblio/bibtex/contrib/doc/
	% The IEEEtran BibTeX style support page is at:
	% http://www.michaelshell.org/tex/ieeetran/bibtex/
	%\bibliographystyle{IEEEtran}
	% argument is your BibTeX string definitions and bibliography database(s)
	%\bibliography{IEEEabrv,../bib/paper}
	%
	% <OR> manually copy in the resultant .bbl file
	% set second argument of \begin to the number of references
	% (used to reserve space for the reference number labels box)
	%\begin{thebibliography}{1}
	%
	%\bibitem{IEEEhowto:kopka}
	%H.~Kopka and P.~W. Daly, \emph{A Guide to \LaTeX}, 3rd~ed.\hskip 1em plus
	%  0.5em minus 0.4em\relax Harlow, England: Addison-Wesley, 1999.
	%
	%\end{thebibliography}
	%
	\vspace{-3mm}
	\bibliographystyle{IEEEtran}
	\bibliography{sigproc}

	% that's all folks
\end{document}